\documentclass[prl,preprint]{revtex4}
\usepackage{graphicx,color}
\usepackage{epstopdf}
 \begin{document}
\title{A Kinetic Theory of Coupled Oscillators}

\author{Eric J. Hildebrand}
\affiliation{Department of Physics and Astronomy, University of
  Pittsburgh, Pittsburgh, PA} 
\author{Michael A. Buice}
\author{Carson C. Chow}
\affiliation{Laboratory of Biological Modeling, NIDDK, NIH, Bethesda, MD}
\date{\today}

\begin{abstract}
We present an approach for the description of
fluctuations that are due to finite system size induced correlations in
the Kuramoto model of coupled oscillators.  We construct a 
hierarchy for the moments of the density of oscillators that is
analogous to the BBGKY 
hierarchy in the kinetic theory of plasmas and gases.  To calculate
the lowest order system size effect, we truncate
this hierarchy at second 
order and solve the resulting closed equations for the two-oscillator
correlation function around the incoherent state.  We use this
correlation function to compute the 
fluctuations of the order parameter, including the effect of
transients, and compare this computation with 
numerical simulations.
\end{abstract} 

\maketitle

Systems of coupled oscillators appear as models for the dynamics of a
wide  range of phenomena
\cite{winfree,liu,golomb,ermentrout84,ermentrout91,walker,wies}.   The
Kuramoto model is a simple and oft-studied description of coupled 
oscillators which, in the limit of an infinite number of oscillators,
exhibits a phase transition from an incoherent state to phase locked
dynamics~\cite{kuramoto,kuramoto2, kuramotoNishikawa2, strogatz}.
However, numerical simulations show the appearance of fluctuations
that are due to finite system size effects even in the absence of any
external noise.  Because the system is deterministic, these
fluctuations are a manifestation of multi-oscillator correlations and are
expected
to vanish  in the infinite oscillator limit, 
with potentially divergent behavior near the transition \cite{daido4}.
While there has been some effort towards an analytic treatment of
the fluctuations in the Kuramoto model~\cite{kuramotoNishikawa, daido3}, 
there is at present no systematic approach.  Here, we
present a statistical formalism which draws upon the kinetic 
theory of plasmas~\cite{ichimaru,nicholson}.   Our methods are
generalizable to any oscillator model.  

The Kuramoto model describes the phase
evolution of $N$ oscillators and is given by 
\begin{equation}
\dot{\theta_i}=\omega_i+\frac{K}{N}\sum_{j=1}^{N}f(\theta_j-\theta_i),\hspace{0.2in}    
i=1,\ldots,N , \label{eq:kur} 
\end{equation}
where $K$ is the coupling strength; the $\omega_i$ are drawn from a
distribution $g(\omega)$, 
assumed to be symmetric and of zero mean. The coupling function
$f(\theta)$ can be any function. In the original Kuramoto model
$f(\theta)=\sin\theta$, which we use for our simulations. 

\begin{figure}
	\scalebox{0.35}{\includegraphics{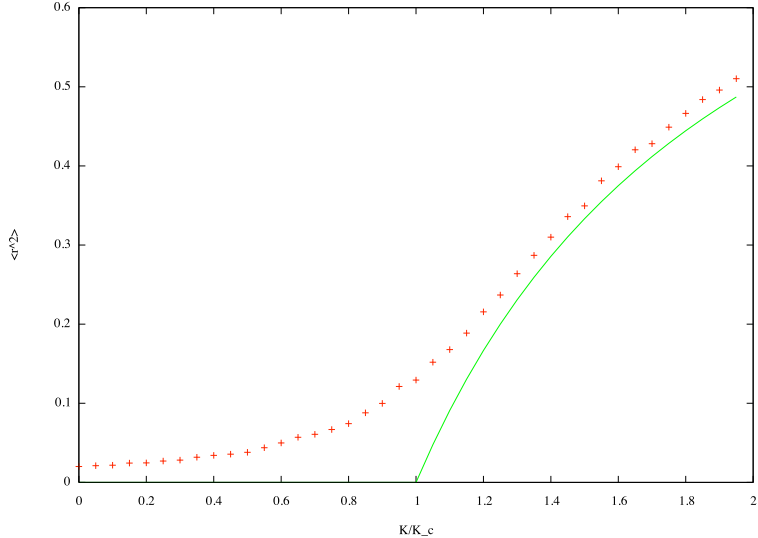}}
	\caption{Phase transition in the Kuramoto model from
	asynchronous ($K<K_c$) to synchronous ($K>K_c$) behavior.  The
	solid line is the mean field prediction for $r^2$; crosses are
	simulation data.} 	\label{fig:phasetransition}
\end{figure}

In the $N\rightarrow
\infty$ limit, Kuramoto showed \cite{kuramoto} that
as the coupling $K$ is increased from $0$, this model exhibits a phase
transition described by the order parameter 
\begin{equation}
Z=\frac{1}{N}\sum_{j=1}^N e^{i\theta_j}\equiv r e^{i\psi}
\label{eq:op_discrete} 
\end{equation}
which is a measure of the level of synchrony in the population.
Kuramoto found a continuous
transition from a phase of complete incoherence
($r= 0$) in the population to a relative degree of
coherence ($r>0$) for $K$ greater
than $K_c=2/\pi g(0)$.  However, for a finite number of oscillators, $r$ will fluctuate.
Figure~\ref{fig:phasetransition} shows $\langle r^2 
\rangle$ (where $\langle \cdot \rangle$ represents an ensemble average
over frequencies and initial angles) for a numerical simulation of
$N=50$ oscillators.  We see 
that the fluctuations smooth the sharp transition from incoherence to
coherence. 

As we will show, typical with phase transitions, the correlations
become enhanced near the onset of the transition (critical point). At low $K$, $\langle r^2 \rangle \approx 1/N$, consistent with the
finite size effects for the free ($K=0$) model. It reasonable to
suppose that in the incoherent state all correlation effects 
(when $K \ne K_c$) are finite size effects due to the
coupling, i.e. the homogeneous all-to-all connectivity forces the
suppression of fluctuations as $N \rightarrow \infty$.   
One of our goals is to calculate $\langle r^2\rangle$ including the
fluctuations due to finite $N$ in the incoherent state.

Mirollo and Strogatz \cite{strogMiro} analyzed the stability of the
incoherent state using a Fokker-Planck formalism.  In the absence of
external additive noise, 
their Fokker-Planck equation has the form of a continuity equation.
They found that the incoherent state has a continuum of marginally
stable modes, which are made stable by additive noise.
In the ensuing, we will generate a series of equations
analogous to the BBGKY hierarchy for which the 
Mirollo-Strogatz continuity equation is the truncation at first
order.  Our strategy is to 
consider an expansion using $1/N$ as a small parameter.   

The complete oscillator probability density
\begin{equation}
n(\theta,\omega,t)=\frac{1}{N}\sum_{i=1}^{N}
\delta(\theta-\theta_i(t))\delta(\omega-\omega_i)
\label{eq:fulldist}
\end{equation}
satisfies the continuity equation
\begin{equation}
\frac{\partial n}{\partial t}+\omega \frac{\partial n}{\partial
  \theta}  
= -K\frac{\partial}{\partial
  \theta}\int_{-\infty}^{\infty}\int_{0}^{2\pi}f(\theta'-\theta)n(\theta',\omega',t)n(\theta,\omega,t)d\theta'
  d\omega'. 
\label{eq:cont} 
\end{equation}
Equation~(\ref{eq:cont}) is 
analogous to the Klimontovich equation in the plasma context and is
still an exact description of the microscopic dynamics.  The complete
oscillator distribution (\ref{eq:fulldist}) 
satisfies the Klimontovich equation~(\ref{eq:cont}) exactly if the
Kuramoto system~(\ref{eq:kur}) is obeyed.
Solving the 
Klimontovich equation for the complete distribution is equivalent to
solving the original system and is equally
difficult.  The strategy of kinetic theory is to consider
the smoothed probability density functions of
the oscillators by taking
ensemble averages.


The one-particle probability density function (PDF) (first moment of
$n(\theta,\omega,t)$) is  given by
\begin{equation}
	\rho_1 (\theta, \omega, t) \equiv \langle n(\theta, \omega, t) \rangle
\end{equation}
where brackets denote the  ensemble average over initial conditions and frequencies. The density $\rho_1 d\theta
d\omega$ represents the mean fraction of oscillators within frequency
range $(\omega,\omega + d\omega)$ and angle range $(\theta, \theta +
d\theta)$.    We note that
$\int_0^{2\pi} \rho_1(\theta,\omega,t) d\theta = g(\omega)$.
Henceforth, we will use the compact notation
$x = (\theta, \omega)$. 
Taking the expectation value of Eq.~(\ref{eq:cont}) gives
\begin{eqnarray}
\frac{\partial \rho_1}{\partial t}+\omega \frac{\partial \rho_1}{\partial \theta}
&+&K\frac{\partial}{\partial \theta}\int_{-\infty}^{\infty}\int_{0}^{2\pi}f(\theta'-\theta)\rho_1(x,t)\rho_1(x',t)d\theta' d\omega' \nonumber\\
&=&-K\frac{\partial}{\partial \theta}\int_{-\infty}^{\infty}\int_{0}^{2\pi}f(\theta'-\theta)C(x;x',t)d\theta' d\omega' \label{eq:bbgky1b}
\end{eqnarray}
where 
\begin{equation}
	C(x;x', t) = \rho_2(x;x',t) - \rho_1(x,t) \rho_1(x',t) \label{eq:connected}
\end{equation}
is the ``connected" two oscillator correlation or moment function
with
\begin{equation}
	\langle n(x, t) n(x', t) \rangle = \rho_2(x; x', t) + \frac{1}{N} \delta(x-x')  \rho_1(x, t). \label{eq:disconnected}
\end{equation}
The self-fluctuation term drops out in Eq.~(\ref{eq:bbgky1b}) because we
consider $f(0)=0$.

The RHS of (\ref{eq:bbgky1b}) describes two oscillator interactions
and is comparable to the collision integral from the kinetic theory of gases and
plasmas.  Neglecting the collision integral leads to the Vlasov
equation,  which amounts to a mean field approximation.  The Vlasov
equation and corresponding Fokker-Planck equation, which includes
a diffusive term when external noise
is included, has been studied for 
coupled oscillators previously in many
contexts~\cite{strogMiro,sakaguchi,treves,abbott, brunel,golomb,cai,cai3}. 
Although the Vlasov equation has the same form as
Eq.~(\ref{eq:cont}), the two should not be confused.
$\rho_1(x,t)$ is a smooth function representing the expectation value
of the number density over initial conditions and frequencies, whereas
$n(x,t)$ is an operator-valued distribution and contains
\emph{all} statistical information about the system. 

We obtain an equation for $C(x;x',t)$ by multiplying
Eq.~(\ref{eq:cont}) by $n(x',t)$ and taking the expectation
value.  This will result in an equation that depends on the three
oscillator moment function.  Continuing this process for higher
moments results in the BBGKY   
hierarchy~\cite{ichimaru,nicholson}.   We 
truncate the hierarchy at second order, expecting
the correlation $C(x;x', t)$ to be $O(1/N)$ and a
general connected $n$-point function to be $O(1/N^{n-1})$ as is
consistent with previous simulations \cite{daido4,daido3}.   

Using Eq.~(\ref{eq:bbgky1b}) and removing terms expected to be
$O(1/N^2)$ yields 
\begin{eqnarray}
\{\frac{\partial}{\partial t}&+&\omega_1\frac{\partial}{\partial \theta_1}+\omega_2 \frac{\partial}{\partial \theta_2}
+K\int_{-\infty}^{\infty}\int_{0}^{2\pi} [\frac{\partial}{\partial \theta_1}f(\theta_3-\theta_1) +\frac{\partial}{\partial \theta_2}f(\theta_3-\theta_2)]\rho_1(x_3,t)d\theta_3 d\omega_3 \}C(x_1,x_2,t)\nonumber \\
&+& K\int_{-\infty}^{\infty}\int_{0}^{2\pi} \frac{\partial}{\partial \theta_1}f(\theta_3-\theta_1)\rho_1(x_1,t)C(x_2,x_3,t)d\theta_3 d\omega_3\nonumber\\
&+& K\int_{-\infty}^{\infty}\int_{0}^{2\pi} \frac{\partial}{\partial \theta_2}f(\theta_3-\theta_2)\rho_1(x_2,t)C(x_3,x_1,t)d\theta_3 d\omega_3 \} \nonumber \\
&=&-\frac{K}{N}[\frac{\partial}{\partial \theta_1}f(\theta_2-\theta_1)
  +\frac{\partial}{\partial
    \theta_2}f(\theta_1-\theta_2)]\rho_1(x_1,t)\rho_1(x_2,t),
\label{eq:C_rho1} 
\end{eqnarray}
Equations~(\ref{eq:bbgky1b}) and (\ref{eq:C_rho1}) form a Gaussian
closure of a kinetic theory describing the Kuramoto model.  We use this to calculate the
fluctuations about the incoherent state.  We start with the {\it ansatz}
\cite{ichimaru}: 
\begin{eqnarray}
C(x_1,x_2,t)&=&\int_{-\infty}^{\infty}d\omega_1'd\omega_2'\int_{0}^{2\pi}d\theta_1'd\theta_2'\int_{t_0}^t dt \chi(x_1',x_2',t') \nonumber \\
&\times& P(x_1,x_1',t-t')P(x_2,x_2',t-t'). \label{eq:C_soln}  
\end{eqnarray}
where the initial conditions are imposed at $t_0$ and $t_0 < t' < t$.
Using Eq.~(\ref{eq:C_soln}) in
Eq.~(\ref{eq:C_rho1}), we obtain the dynamics for the propagator $P$,
\begin{eqnarray}
\{\frac{\partial}{\partial t}&+&\omega_1 \frac{\partial}{\partial \theta_1} 
+K\frac{\partial}{\partial \theta_1}\int_{-\infty}^{\infty}\int_{0}^{2\pi}f(\theta_2-\theta_1)\rho_1(x_2,t)d\theta_2 d\omega_2\}P(x_1,x_1',t-t') \nonumber \\
&+&K\frac{\partial}{\partial \theta_1}\int_{-\infty}^{\infty}\int_{0}^{2\pi}f(\theta_2-\theta_1)\rho_1(x_1,t)P(x_2,x_1',t-t')d\theta_2 d\omega_2=0, 
\label{eq:prop}
\end{eqnarray}
where
\begin{equation}
\chi(x_1,x_2,t)=
-\frac{K}{N}[\frac{\partial}{\partial \theta_1}f(\theta_2-\theta_1)
  +\frac{\partial}{\partial
    \theta_2}f(\theta_1-\theta_2)]\rho_1(x_1,t)\rho_1(x_2,t)
\label{eq:C_0}  
\end{equation}
and the initial condition is $P(x;x', 0) = \delta(x-x')$.
 
The fluctuations in the order parameter
are given by 
\begin{equation}
\langle r^2 \rangle \equiv \langle Z Z^* \rangle =  \int d\omega d\omega' d\theta d\theta' \langle n(\omega,\theta, t) n( \omega', \theta',t)\rangle e^{i(\theta - \theta')} 
\label{eq:flucs}
\end{equation}
We consider fluctuations in the incoherent state and thus seek
solutions to (\ref{eq:bbgky1b}) and (\ref{eq:C_rho1}) such that
$\rho_1(x,t) = \frac{1}{2\pi}g(\omega)$.   From
Eqs. (\ref{eq:connected}) and (\ref{eq:disconnected}), we see that a
computation of the fluctuations amounts to a calculation of the
connected correlation function, which is phase invariant (because
$\rho_1$ is independent of $\theta$), so that
$C(\theta_1,\theta_2,\omega_1,\omega_2,t)=C(\theta_1-\theta_2,\omega_1,\omega_2,t)$. 
Hence, the collision integral in Eq.~(\ref{eq:bbgky1b})
is zero, making $\rho_1(\theta, \omega) = g(\omega)/2\pi$ an exact
solution of the equations. 
Taking the Fourier and Laplace transforms in $\theta$ and time of
Eq.~(\ref{eq:C_soln}) gives
\begin{eqnarray}
C_n(\omega_1,\omega_2, s) &=&
\frac{nK{\rm
    Im}[f_n]}{2\pi^2N}\int_{-\infty}^{\infty}d\omega_1'd\omega_2'
d\tau\int_{{\cal L}_1} ds_1 \int_{{\cal L}_2} ds_2 g(\omega'_1)
g(\omega'_2) \label{eq:CLT}  \\ 
&\times& \hat{P}_{-n}(\omega_1,\omega_1',s_1)\hat{P}_n(\omega_2,\omega_2',s_2)\frac{1}{s} e^{(s_1+s_2 - s)\tau},  \nonumber
\end{eqnarray}
where $n$ is the Fourier mode index, $s_{1,2}$ is a Laplace transform
variable and $\tau=t-t'$.
By definition, the Laplace contours ${\cal L}_1$ and ${\cal L}_2$ are
arranged such that they are to the right of all poles in
$\hat{P}_{-n}$ and $\hat{P}_{n}$, respectively.  
Using  Eqs. (\ref{eq:connected}), (\ref{eq:disconnected}) and (\ref{eq:CLT}) in Eq.~(\ref{eq:flucs}) gives
\begin{equation}
\langle r^2(\tau) \rangle = 4\pi^2 \int d\omega d\omega' C_{-1}(\omega, \omega', \tau) + \frac{1}{N}
\label{eq:r2gen}
\end{equation}
because $\langle Z \rangle=0$. 

We can obtain a general expression for $\langle r^2 \rangle$ 
without explicitly solving for the correlation
function.  From  equation~(\ref{eq:prop}), we can derive
the relation
\begin{equation}
	\int \hat{P}_n(\omega, \omega', s) d\omega = \frac{1}{(s +
	in\omega')\Lambda_n(s)} \label{eq:P}
\end{equation}
where
\begin{equation}
\Lambda_n(s) \equiv 1+inKf_n^{\ast}\int_{-\infty}^{\infty}\frac{g(\omega) d\omega}{s + in\omega} \label{eq:Lambda0}
\end{equation}
is the analog of the
dielectric response function. 
Using Eqs.~(\ref{eq:P}) and  (\ref{eq:CLT}) in Eq.~(\ref{eq:r2gen}) yields
\begin{equation}
	\langle r^2(\tau) \rangle = \frac{2}{iKN\pi} \int_{{\cal C}} ds \frac{\Lambda_1(s - s_0) - 1}{\Lambda_1(s-s_0)} {\rm Res}\left [ \frac{-1}{\Lambda_1(s)}\right ]_{s = s_0} \frac{1}{s}e^{s\tau}
	\label{eq:r2spec}
\end{equation}
where $s_0$ is the zero of $\Lambda_n(s)$.  The strategy of the
calculation leading to Eq.~(\ref{eq:r2spec}) is similar to the
calculation of the Lenard-Balescu collision
integral~\cite{ichimaru,nicholson}.

For the specific frequency distribution
$g(\omega)=(\gamma/\pi)(1/(\omega^2+\gamma^2))$ (i.e. a Lorentz
distribution), Eq.~(\ref{eq:r2spec})
evaluates to
\begin{equation}
	\langle r^2(\tau) \rangle = \frac{1}{N} \frac{K_c}{K_c - K} - \frac{1}{N} \frac{K}{K_c - K} e^{-( K_c - K)\tau} \label{eq:r2}
\end{equation}
where $K_c=2\gamma$ for the Lorentz frequency distribution.   $\langle
r^2(0) \rangle =1/N$  because the initial conditions for
Eqs.~(\ref{eq:bbgky1b}) and (\ref{eq:C_rho1}) are such that
$\rho_1(x,0)$ is the equilibrium incoherent state and $C(x_1, x_2, 0)
= 0$. 
For the uncoupled system $K=0$, so $\langle
r^2\rangle=1/N$ as expected.  We also see that the amplitude of the
fluctuations and the transient decay time become 
singular at the critical point $K=K_c$.  At criticality, we obtain the
expression 
$\langle r^2(\tau) \rangle = (1/N) (1+K_c \tau)$.  The
closer $K$ is to criticality, the less this calculation should be
valid.  Near critical behavior requires an analysis of all orders in
the $1/N$ expansion.  Dynamically, the implication is that as the
coupling strength nears criticality, oscillators will interact more
strongly and higher order correlations will become more important. 
The result $\langle r^2(\infty) \rangle =
K_c/[N(K_c - K)]$ was first derived by Daido
\cite{daido3} with a completely different approach.  Our method facilitates
a systematic expansion  in $1/N$, in
addition to providing an examination of the transient behavior of
$\langle r^2(\tau) \rangle$.

We can examine the transient behavior of the correlations by solving
Eq.~(\ref{eq:CLT}) for the Lorentz distribution. We first solve for the propagator
in Eq.~(\ref{eq:prop}) by taking a Fourier series
expansion in $\theta$ and Laplace transform in time, to obtain 
\begin{equation}
\hat{P}_n(\omega_1,\omega_1',s)=\frac{1}{s+in\omega_1}
  \frac{\delta(\omega_1-\omega_1')}{2\pi} -\frac{inKf_n^\ast 
  g(\omega_1)}{2\pi(s+in\omega_1)(s+in\omega_1')\Lambda_n(s)} 
\label{eq:prop_soln}
\end{equation}
where $s$ is the Laplace transform variable and
$\Lambda_n(s)  = 1 -(K/2)|n|/(s + |n|\gamma)$.
The propagator (\ref{eq:prop_soln}) has poles along the imaginary axis
corresponding to 
the continuous spectrum of marginally stable modes
as well as  those given by the discrete zeros of $\Lambda_n(s)$,
which for
$K<K_c = 2\gamma$ are real and negative~\cite{strogMiro}.

We then 
use Eq.~(\ref{eq:prop_soln}) in Eq.~(\ref{eq:CLT}) and take the
inverse Laplace transform.  For the $n=1$ mode, this gives
\begin{eqnarray}
	C_{1}(\omega, \omega', \tau)&=&  \frac{K}{N}\frac{1}{4\pi^2}g(\omega)g(\omega')   \left [  \frac{(i\omega + \frac{K_c}{2})(-i\omega' + \frac{K_c}{2})}{(i\omega + \frac{K_c}{2} - \frac{K}{2})(-i\omega' + \frac{K_c}{2} - \frac{K}{2})} \left (\frac{e^{(i(\omega  - \omega') \tau)}}{i(\omega - \omega')} - \frac{1}{i(\omega - \omega')} \right) \right . \nonumber \\
	&+& \frac{K}{2} \frac{ (  i \omega' - \frac{K_c}{2})}{(\frac{K}{2} - \frac{K_c}{2} - i \omega)((\frac{K_c}{2} - \frac{K}{2})^2 + (\omega')^2)} \left ( e^{-(\frac{K_c}{2} - \frac{K}{2} +i \omega')\tau} - 1\right) \nonumber \\
	&+& \frac{K}{2} \frac{ (  -i \omega - \frac{K_c}{2})}{(\frac{K}{2} - \frac{K_c}{2} + i \omega')((\frac{K_c}{2} - \frac{K}{2})^2 + (\omega)^2)} \left ( e^{-(\frac{K_c}{2} - \frac{K}{2} -i \omega)\tau} - 1\right) \nonumber \\
	&+&\left .  \frac{K^2}{4(K-2\frac{K_c}{2})} \frac{1}{(\frac{K}{2} - \frac{K_c}{2} - i \omega)(\frac{K}{2} - \frac{K_c}{2} + i \omega')} \left ( e^{-(K_c - K)\tau} -1 \right)\right ] \label{eq:cor}
\end{eqnarray}
where $\tau = t - t_0$.  The other modes are given by $C_{-1} = C_{1}^*$ and $C_n = 0$ for $n \ne \pm 1$ since $f(\theta)=\sin\theta$.
The correlation function will thus have the form
\begin{equation}
	C(\omega, \omega', \theta - \theta', \tau) = C_1 e^{i(\theta - \theta')} + C_{-1} e^{-i(\theta - \theta')}
	\label{eq:cForm}
\end{equation}
Only the last term in Eq.~(\ref{eq:cor}) contributes to the transient
in Eq.~(\ref{eq:r2}).  The correlation function contains
modes which consist of all possible pairings of a marginal
oscillating mode with the decaying mode.  While
the marginal modes do not decay in the correlations, they do not
affect the decay of $\langle Z(\tau) \rangle$  because of a
Landau damping-like dephasing effect that is described in
Ref.~\cite{strogMiroMatt}.  The marginal modes also have no effect upon the
transient behavior of $\langle r^2(\tau) \rangle$.   We should expect a similar
result for higher moments. 
At this order, the
marginal modes are not rendered stable by finite size effects as they
are with the addition 
of external additive noise \cite{strogMiro}.  Should stabilization
occur due to the intrinsic fluctuations, it will necessarily be a
consequence of higher order effects.

\begin{figure}
	\scalebox{0.3}{\includegraphics{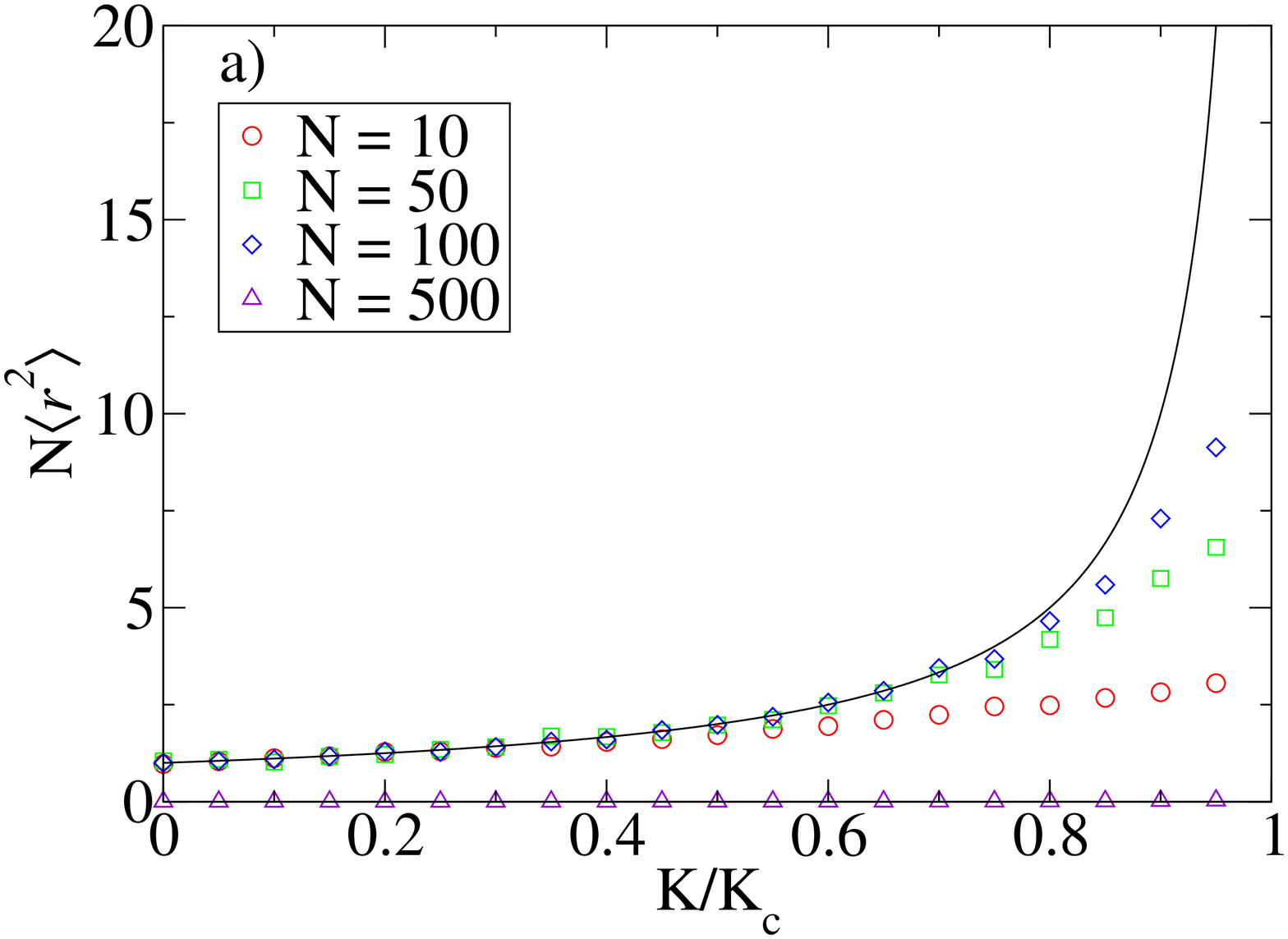}}
	\scalebox{0.3}{\includegraphics{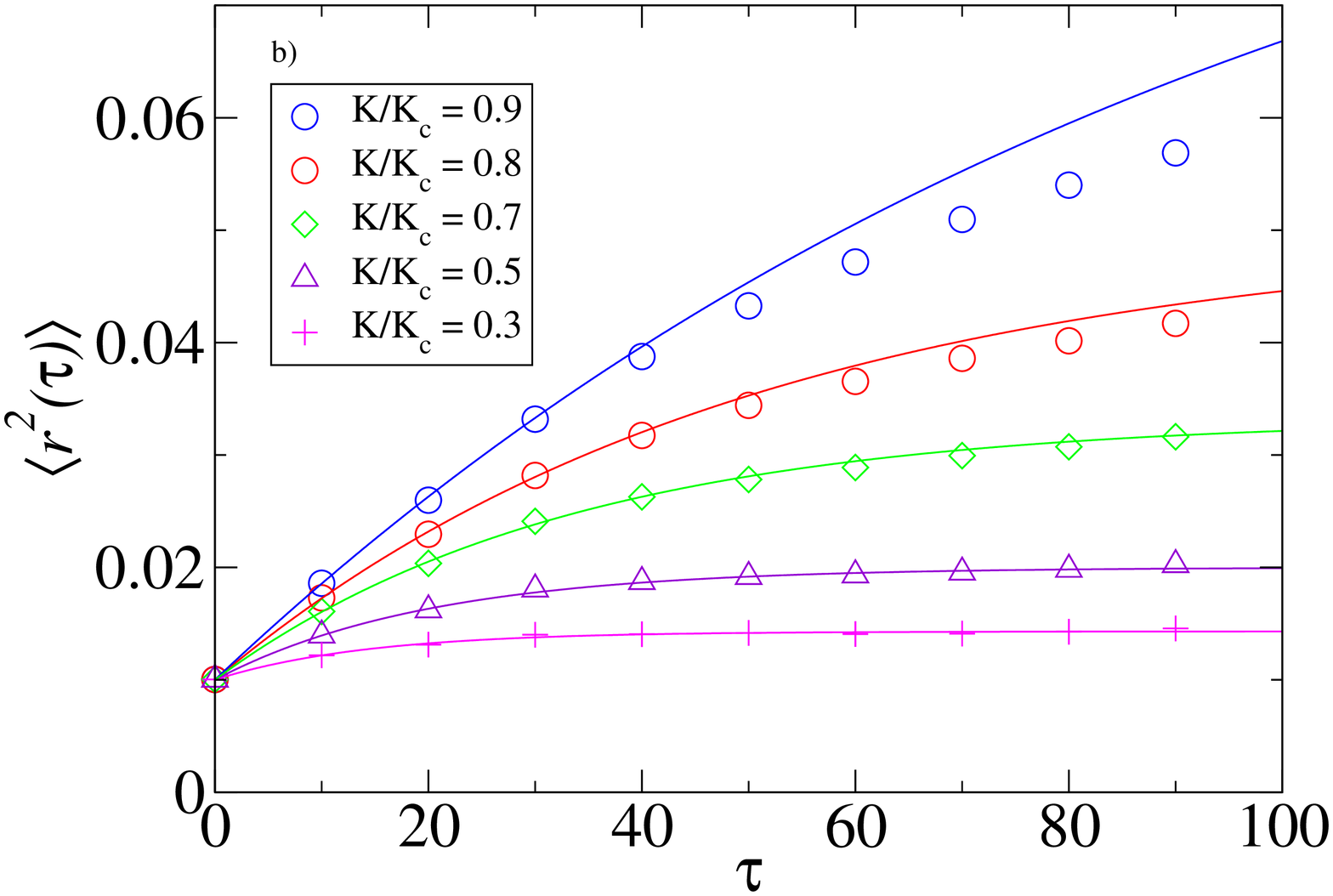}}
	\caption{a) Simulated and predicted $N\langle r^2 \rangle$
	  vs. $K/K_c$ for various values of N for large times.  
The data are evolved for 6 time constants at each $K$ ($\tau = 1/(K_c - K)$) and averaged over 1000 different initial conditions and frequencies.  The frequency distribution is Lorentz with $\gamma = 0.05$.  The simulation was performed with a time step of 0.05.  The initial distribution of angles was uniform. 
b) Time evolution of $\langle r^2(\tau) \rangle$ vs. $\tau$ for various values of $K$ and for $N=100$.  At each time point the values are averaged over 10,000 different initial conditions and frequencies.  All other parameters as above.}
	\label{fig:r2}
\end{figure}
\begin{figure}
	\scalebox{0.35}{\includegraphics{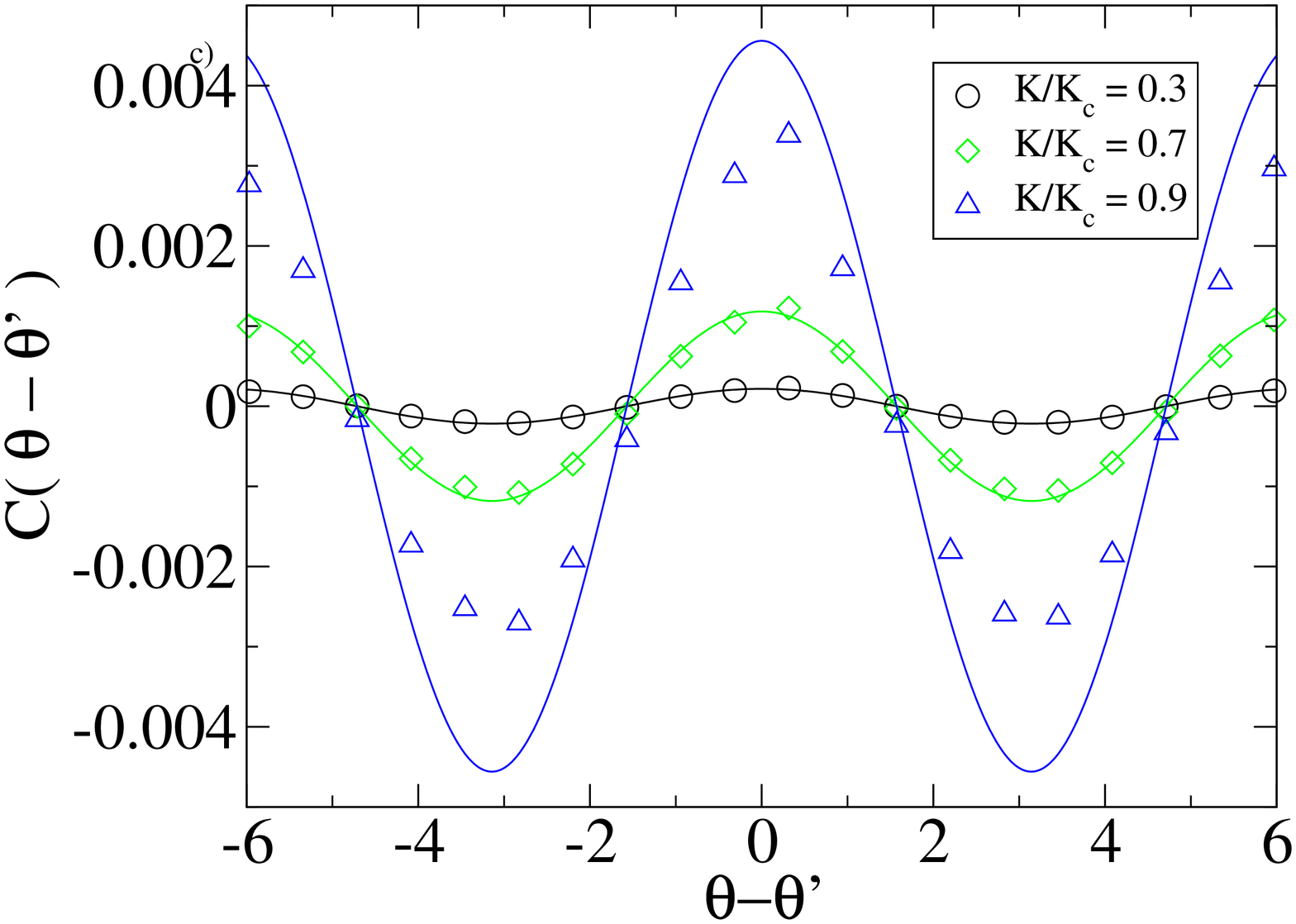}}
	\caption{$C(\omega,\omega',\theta-\theta')$ integrated over
	$\omega$ and $\omega'$ versus $\theta - \theta'$ for $N=100$
	and various values of $K$.  Frequency distribution is Lorentz
	with $\gamma = 0.05$ and the initial angle distribution is
	uniform.   Results are averaged over 100,000 samples in a 2
	dimensional histogram with 100x100 bins.  The data is then
	averaged over angle differences and then put into a one
	dimensional histogram with bins of width 10.  The time step for the evolution is 0.05.}

	\label{fig:cor}
\end{figure}

We compare our analytical results to numerical simulations of the Kuramoto system. Figure~\ref{fig:r2} a) shows the asymptotic value of $\langle r^2\rangle$ for various values of $K$ and $N$. 
 The analytical prediction matches extremely well for $N=500$ and reasonably well for 
 $N=50$ and $N=100$.  Only at $N=10$ are there
significant deviations from  the prediction.    Fig.~\ref{fig:r2} b) shows the transient behavior of $\langle r^2\rangle$.
The results
 match quite well below $K/K_c = 0.8$.  
Numerical results
for  the correlation 
function integrated over $\omega, \omega'$  are shown in Fig.~\ref{fig:cor}.  The simulation agrees well
with the prediction Eq.~(\ref{eq:cForm}) except near the critical
 point as expected.

Our calculation is the first presentation of a systematic approach to
understanding the fluctuations due to finite size effects to an
arbitrary order in $1/N$.  
Although we truncate at lowest order, our
approach allows a truncation at any level of the moment hierarchy to
produce an expansion in $1/N$.  We note that
Ref.~\cite{pikovsky} found that when the oscillators are driven with
Gaussian noise, $1/N$ dependence is 
still seen in the fluctuations of the order parameter.  Additionally,
our methodology could be used to study the evolution of the phase of $Z$, as
in Ref.~\cite{gleeson}. 

Some previous work
~\cite{brunel,golomb, treves, abbott,crawford3, pikovsky} for both
phase and pulse 
coupled oscillators also start with a continuity equation similar to Eq.~(\ref{eq:cont})
but either go directly to mean field theory, with and without
an external noise source to approximate fluctuations, or
assume the fluctuations are Gaussian. 
References~\cite{cai, cai3} derive a kinetic theory for a network of
integrate-and-fire neurons by
constructing a moment hierarchy similar to ours
that is closed using the maximum
entropy principle. However, this work differs from ours in that the
hierarchy is built from a Boltzmann-like equation for a one-particle
distribution function with stochastic inputs and hence does not capture
the same correlation effects that we find by starting
from a continuity equation that contains the full statistics of the
system.

We feel it is
important to 
stress that the Klimontovich continuity equation (Eq.~(\ref{eq:cont})) is not an approximation.  The
approximation appears in the method of finding solutions. The mean
field limit is equivalent to setting correlations to zero.
Computing the moment hierarchy allows for an expansion which
accounts for the effects of correlations.  We produce a
systematic method for deriving such an expansion and show explicitly
in what regime higher order correlations can be ignored.   
We also note that the kinetic theory approach is only one avenue to
understanding correlations.  An alternative formulation is through a
statistical field theory approach \cite{zinnjustin}, which facilitates
the construction of an expansion without resorting to a moment
hierarchy.

This research was supported in part by the Intramural Research Program
of NIH, NIDDK.  

\bibliography{kineticRefs}



%


\end{document}